\documentclass[10pt]{mpi08} \usepackage{graphicx}
\usepackage{epsfig}\usepackage{amssymb}
\usepackage{cite,./mcite}  
\begin{document} \title{Multiple partonic interactions in heavy-ion collisions} 
\author{Cyrille Marquet}
\institute{Institut de Physique Th\'eorique, CEA/Saclay, 91191 Gif-sur-Yvette 
cedex, France\\
Department of Physics, Columbia University, New York, NY 10027, USA}
\maketitle
\begin{abstract}
I discuss the role played by multiple partonic interactions (MPI) in the early stages of relativistic heavy-ion collisions, for which a weak-coupling QCD description is possible. From the Color Glass Condensate, through the Glasma and into the
Quark-Gluon-Plasma phase, MPI are at the origin of interesting novel QCD phenomena.
\end{abstract}

\section{Introduction}

Relativistic heavy-ion collisions involve such large parton densities, that they are
reactions where multiple partonic interactions (MPI) abound, and in which those
can be investigated. Through most of the stages of a high-energy heavy-ion 
collision, MPI are not only important but crucial, and without their understanding,
no robust QCD-based description of the collision can be achieved. During the
different phases that the system goes through, from the initial nuclear wave
functions, through the pre-equilibrium state just after the collision, and into the
following thermalized quark-gluon plasma (QGP) and hadronic phases, MPI are
at the origin of most interesting phenomena.

However, one may wonder what can be described with first-principle weak-coupling
QCD calculations. It has been proposed that the early stages of the heavy-ion
collision should be, perhaps until the QGP phase. The saturation of the initial nuclear wave functions, and the multiparticle production from the decay of strong color fields are phenomena which have been addressed by weak-coupling methods, as well as the quenching of hard probes via QGP-induced energy loss. In those calculations, MPI are characterized by momentum scales which, if hard enough, justify a weak coupling analysis.

In the Color Glass Condensate (CGC) picture of the nuclear wave function, the
saturation scale $Q_s$ characterizes which quantum fluctuations can be treated
incoherently and which cannot; in the glasma phase right after the collision of two
CGCs, $1/Q_s$ sets the time scale for the decay of the strong color fields; and in the QGP phase, the plasma saturation momentum characterizes what part of the wave
function of hard probes is responsible for their energy loss, by becoming emitted
radiation. In the following, I discuss the role played by MPI in those different stages.

\section{The saturation scale in the nuclear wave function}

The QCD description of hadrons/nuclei in terms of quarks and gluons depends on
the process under consideration, on what part of the wave function is being probed.
Consider a hadron moving at nearly the speed of light along the light cone direction
$x^+,$ with momentum $P^+.$ Depending on their transverse momentum $k_T$
and longitudinal momentum $xP^+,$ the partons inside the hadron behave differently,
reflecting the different regimes of the hadron wave function.

When probing the (non-perturbative) soft part of the wave function, corresponding 
to partons with transverse momenta of the order of $\Lambda_{QCD}\!\sim\!200\ 
\mbox{MeV},$ the hadron looks like a bound state of strongly interacting partons. 
When probing the hard part of the wave function, corresponding to partons with 
$k_T\!\gg\!\Lambda_{QCD}$ and $x\!\lesssim\!1,$ the hadron looks like a dilute 
system of weakly interacting partons.

\begin{figure}[t]
\begin{minipage}[t]{50mm}
\centerline{\epsfxsize=5cm\epsfbox{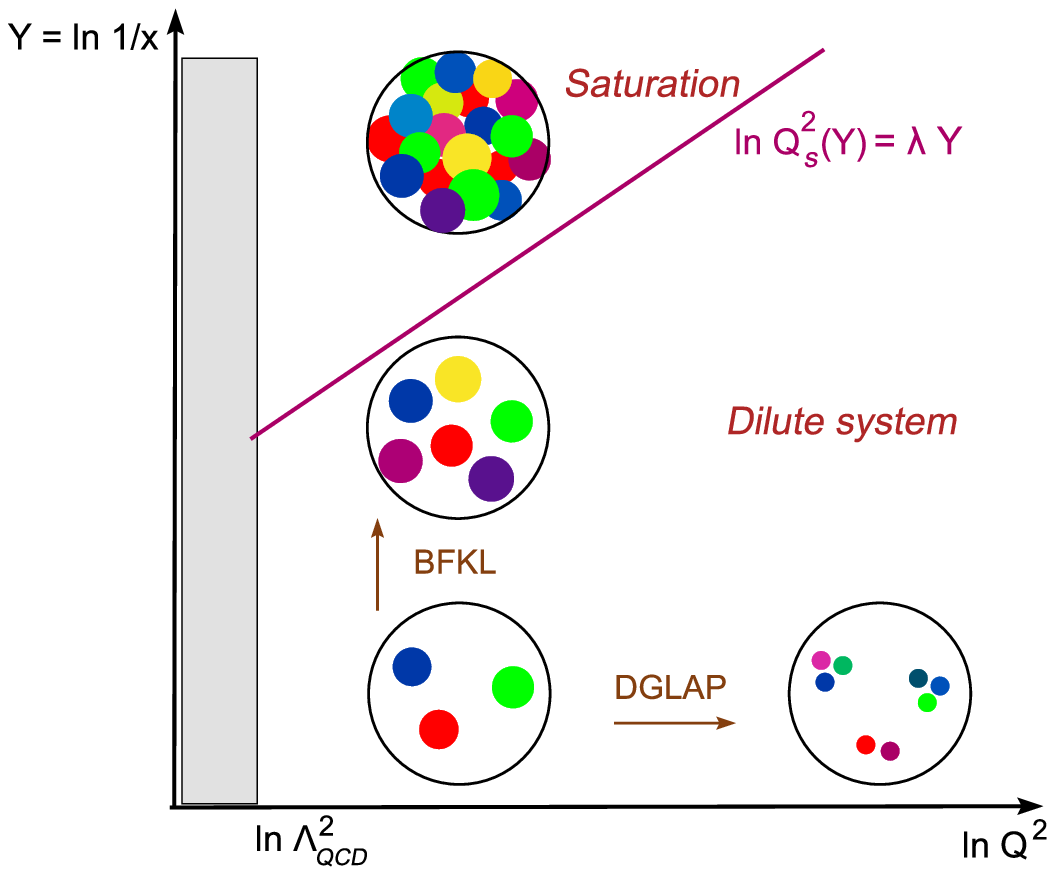}}
\end{minipage}
\hspace{\fill}
\begin{minipage}[t]{80mm}
\vspace{-4cm}\centerline{\epsfxsize=8cm\epsfbox{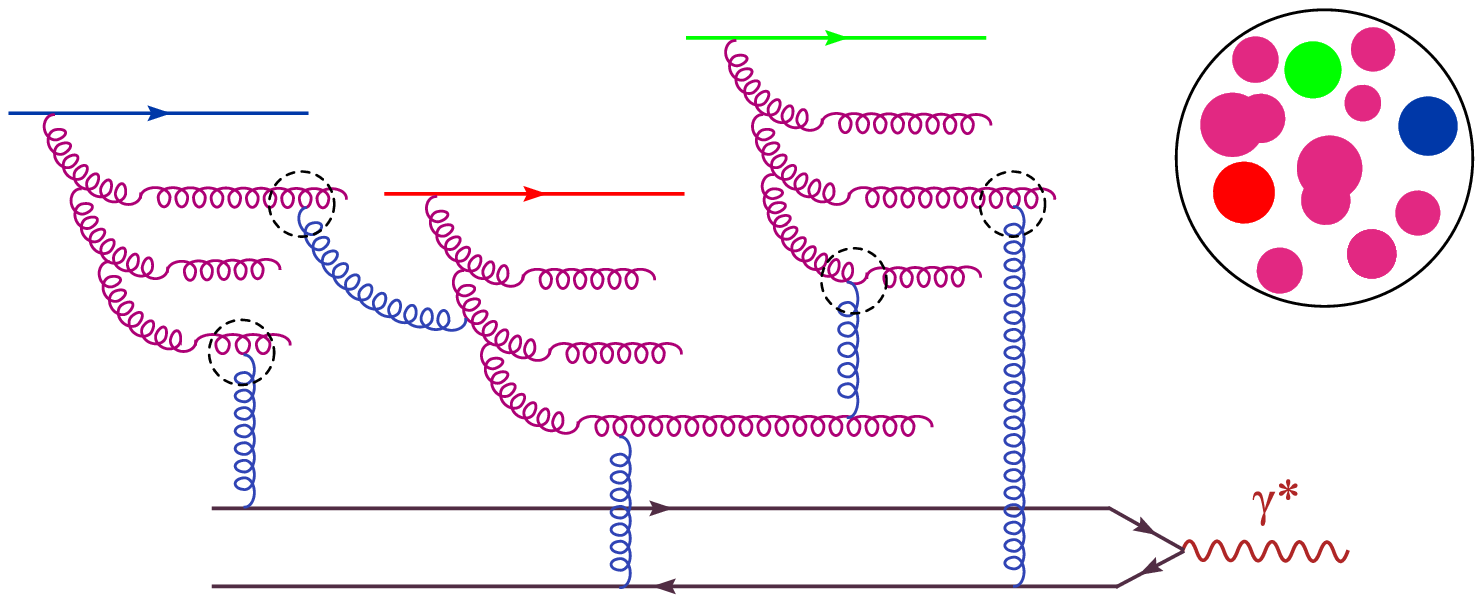}}
\end{minipage}
\caption{Left: diagram in the $(k_T^2\!=\!Q^2,x)$ plane picturing the hadron/nucleus in the different weakly-coupled regimes. The saturation line separates the dilute (leading-twist) regime from the dense (saturation) regime. Right: when scattering a dilute probe on the hadron/nucleus, both multiple scatterings and saturation of the wave function are equally important at small $x,$ when occupation numbers become of order $1/\alpha_s.$}
\end{figure}

The saturation regime of QCD describes the small$-x$ part of the wave function.
When probing partons that feature $k_T\!\gg\!\Lambda_{QCD},$ and $x\!\ll\!1,$
the effective coupling constant $\alpha_s\log(1/x)$ is large, and the hadron looks
like a dense system of weakly interacting partons, mainly gluons (called small$-x$ gluons).
The larger $k_T$ is, the smallest $x$ needs to be to enter the saturation 
regime. As pictured in Fig.1, this means that the separation between the dense 
and dilute regimes is characterized by a momentum scale $Q_s(x),$ called the 
saturation scale, which increases as $x$ decreases. 

A simple way to estimate the saturation scale is to equate the gluon-recombination cross-section
$\sigma_{rec}\sim\alpha_s/k_T^2$ with $1/\rho_T\sim \pi R^2/(xf(x,k_T^2)),$ the inverse gluon
density per unit of transverse area. Indeed, when $\sigma_{rec}\rho_T\sim 1,$ one expects
recombination not to be negligible anymore. This gives:
\begin{equation}
Q_s^2=\frac{\alpha_s xf(x,Q_s^2)}{\pi R^2}\ .
\end{equation}
Note that $\alpha_s(Q_s^2)$ decreases as $x$ decreases, so for small enough $x,$ one
deals with a weakly-coupled regime, even though non-linear effects are important. The scattering
of dilute partons (with $k_T\!\gg\!Q_s(x)$) is described in the leading-twist approximation in
which they scatter incoherently. By contrast, when the parton density is large $(k_T\!\sim\!Q_s(x)),$
partons scatter collectively.

The Color Glass Condensate (CGC) is an effective theory of QCD \cite{cgcrev} 
which aims at describing this part of the wave function. Rather than using a 
standard Fock-state decomposition, it is more efficient to describe it with 
collective degrees of freedom, more adapted to account for the collective 
behavior of the small-$x$ gluons. The CGC approach uses classical color fields: 
\begin{equation}
|h\rangle=|qqq\rangle+|qqqg\rangle+\dots+|qqqg\dots ggg\rangle+\dots\quad
\Rightarrow\quad|h\rangle=\int D\rho\ \Phi_{x_A}[\rho]\ |\rho\rangle
\label{cgc}\ .\end{equation}
The long-lived, large-$x$ partons are represented by a strong color source 
$\rho\!\sim\!1/g_S$ which is static during the lifetime of the short-lived 
small-$x$ gluons, whose dynamics is described by the color field ${\cal 
A}\!\sim\!1/g_S.$ The arbitrary separation between the field and the source is 
denoted $x_A.$ When probing the CGC with a dilute object carrying a weak color
charge, the color field ${\cal A}$ is directly obtained from $\rho$ via classical Yang-Mills equations:
\begin{equation}
[D_\mu,F^{\mu\nu}]=\delta^{+\nu}\rho\ ,
\end{equation}
and it can be used to characterize the CGC wave function $\Phi_{x_A}[{\cal A}].$

\begin{figure}[t]
\begin{center}
\epsfig{file=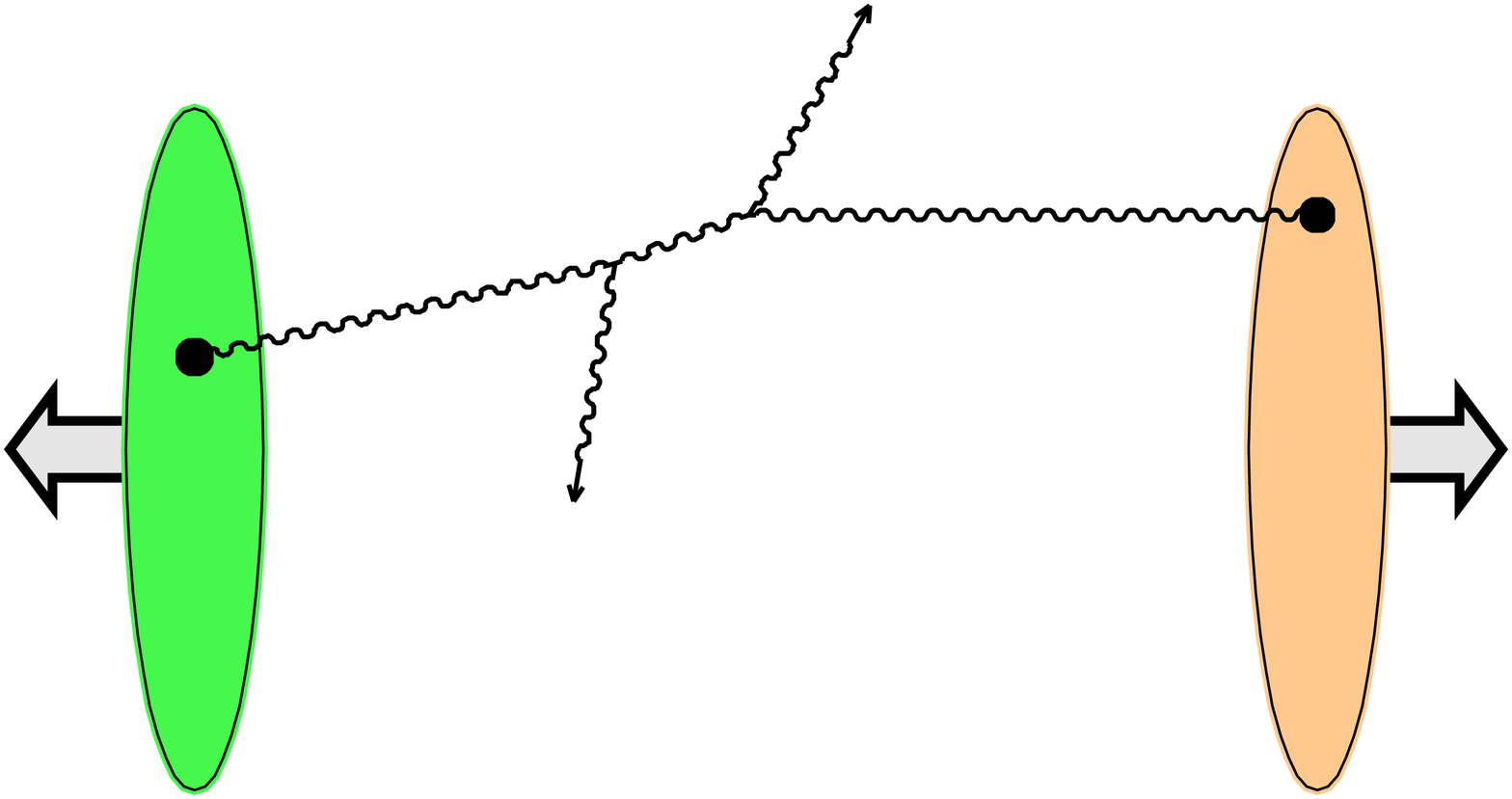,width=6cm}
\hspace{1cm}
\epsfig{file=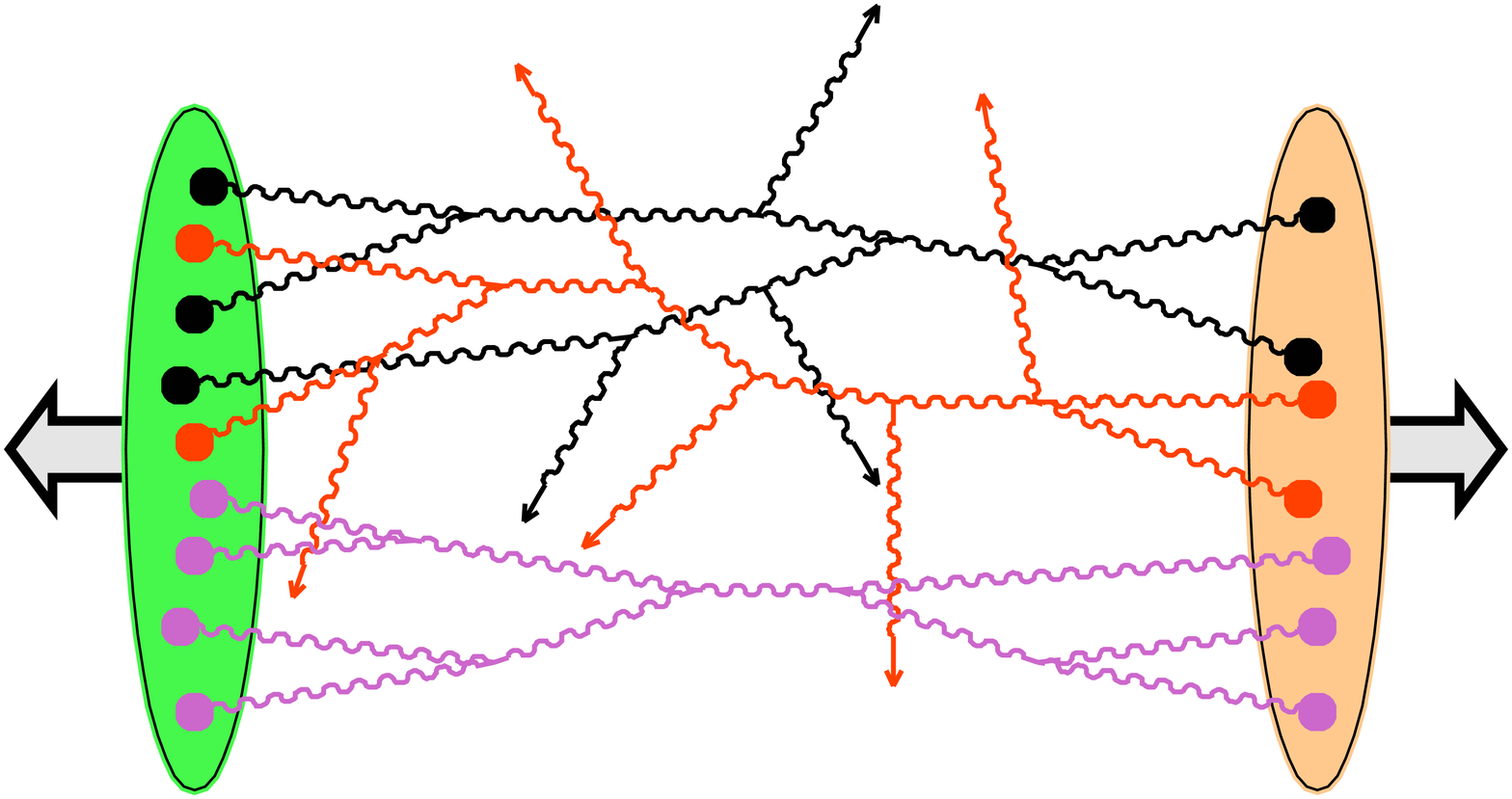,width=6cm}
\caption{Left: typical diagram for the production of high$-p_T$ particles, with large values of $x$ being probed in the nuclear wave functions. Right: typical diagram for the production of bulk particles with $p_T\sim Q_s,$ where multiple partonic interactions are crucial. This is true in heavy-ion collisions, and pp collisions at very high energies.}
\end{center}
\end{figure}

This wave function is a fundamental object of this picture, it is mainly a non-perturbative quantity, but the $x_A$ evolution can be computed perturbatively. Requiring that observables are independent of the choice of $x_A,$ a functional renormalization group equation can be derived. In the leading-logarithmic approximation which resums powers of $\alpha_S\ln(1/x_A),$ the JIMWLK equation describes the evolution of $|\Phi_{x_A}[{\cal A}]|^2$ with $x_A.$ The evolution of the saturation scale with $x$ is then obtained from this equation.

Finally, the information contained in the wave function, on gluon number and gluon 
correlations, can be expressed in terms of n-point correlators, probed in scattering processes.
These correlators consist of Wilson lines averaged with the CGC wave function, and resum powers
of $g_S{\cal A}\sim1,$ {\it i.e.} scattering with an arbitrary number of gluons exchanged.
In the CGC picture, both multiple scatterings and non-linear QCD evolution are taken into account.
Note that in terms of occupation numbers, in the saturation regime one reaches
\begin{equation}
\langle{\cal AA}\rangle=\int D{\cal A}\ |\Phi_{x_A}[{\cal A}]|^2{\cal AA}\sim1/\alpha_s\ .
\end{equation}
Therefore, taking into account multiple scatterings in the collision is as important as the saturation
of the wave function. A consistent calculation of MPI must include both.

It was not obvious that the CGC picture (\ref{cgc}), which requires small values of $x_A,$ would be relevant at present energies. One of the most acclaimed successes came in the context of d+Au collisions at RHIC, where forward particle production $pA\!\to\!hX$  allows to reach small values of $x_A$ with a dilute probe well understood in QCD \cite{jyrev}. The prediction that the yield of high-$p_T$ particles at forward rapidities in pA collisions is suppressed compared to $A$ pp collisions, and should decrease when increasing the rapidity, was confirmed.

\section{Multiple partonic interactions in the Glasma}

The Glasma is the result of the collision of two CGCs. In a high-energy heavy-ion collision, each nuclear
wave function is characterized by a strong color charge, and the field describing the dynamics of the
small-x gluons is the solution of
\begin{equation}
[D_\mu,F^{\mu\nu}]=\delta^{+\nu}\rho_1+\delta^{-\nu}\rho_2\ .
\end{equation}
The field after the collision is non-trivial \cite{glasma}: it has a strong component ($A^\mu\sim1/g_s$), a component
which is particle like ($A^\mu\sim1$), and components of any strength in between.
To understand how this pre-equilibrium system thermalizes, one needs to understand how the
Glasma field decays into particles. Right after the collision, the strong field component contains all modes.
Then, as the field decays, modes with $p_T>1/\tau$ are not part of the strong component
anymore, and for those a particle description becomes more appropriate. After a time of order $1/Q_s,$
this picture breaks down, and it has been a formidable challenge to determine weather a fast thermalization
can be achieved within this framework, due to instabilities \cite{paulraju,*hirokazu}.

A problem which can be more easily addressed is multiparticle production. The difficult task is to express the
cross-section in terms of the Glasma field, and this is when MPI must be dealt with, as pictured in Fig.2.
This has first been done at tree level, and from the one-loop calculation a factorization theorem could then be derived
\cite{glv1,*glv2} (note an interesting possible application of the results to pp collisions:
those first-principle calculations could inspire a model for the underlying event).
Predictions for the total charged-particle multiplicity in AA collisions at the LHC are shown in Fig.3.
Two approaches are compared: in the first, a simplified factorization (called $k_T$ factorization) is
assumed but the energy evolution is accurately obtained from a next-to-leading evolution equation \cite{javier}; in the second, the energy evolution is only parameterized but MPI are correctly dealt with by solving classical Yang-Mills equations \cite{tuomas}. While a full next-leading treatment of both multiple scatterings and small-$x$ evolution is desirable, the numbers obtained are similar, which indicates that the uncertainties in both approaches are under control. 

\begin{figure}[t]
\begin{center}
\epsfig{file=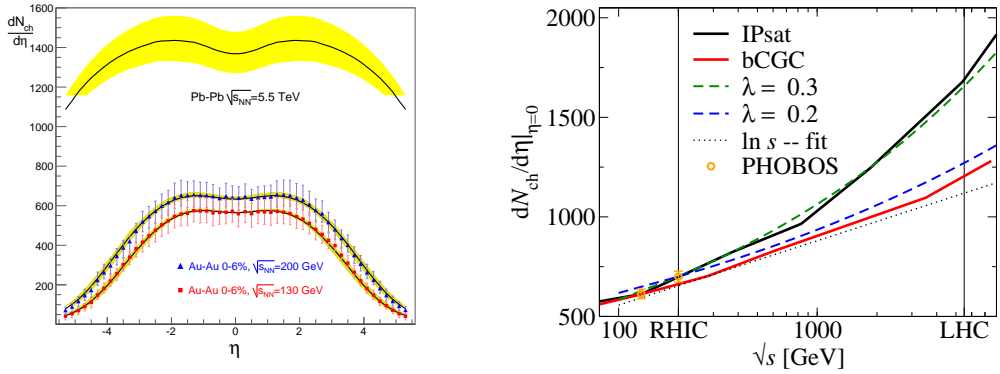,width=5.5cm}
\hspace{1cm}
\epsfig{file=tuomas.eps,width=6.5cm}
\caption{The charged-particle multiplicity in AA collisions at RHIC and the LHC. 
In both approaches a few parameters are fixed to reproduce RHIC data, such as 
the initial value of $Q_s.$ Then the small-$x$ evolution determines the multiplicity at the LHC. The predictions are similar, around 1400 charged particles at mid rapidity for central collisions.}
\end{center}
\end{figure}

\section{The saturation scale in the QCD plasma}

Hard probes are believed to be understood well enough to provide clean measurements of the properties
of the QGP formed in heavy-ion collisions. A large amount of work has been devoted to understand
what happens to a quark (of high energy $E,$ mass $M$ and Lorentz factor  $\gamma=E/M$) as it propagates
through a thermalized plasma \cite{reviewKW,*reviewdde}. MPI are a main ingredient of the perturbative QCD (pQCD)
description of how a quark losses energy, until it thermalizes or exits the medium (see Fig.4).

At lowest order with respect to $\alpha_s,$ quantum fluctuations in a quark wave function consist of a single
gluon, whose energy we denote $\omega$ and transverse momentum $k_\perp.$ The virtuality of that
fluctuation is measured by the coherence time, or lifetime, of the gluon $t_c=\omega/k_\perp^2.$
Short-lived fluctuations are highly virtual while longer-lived fluctuations are more easily put on shell when
they interact. The probability of the fluctuation is $\alpha_sN_c,$ up to a kinematic factor which for heavy
quarks suppresses fluctuations with $\omega>\gamma k_\perp.$ This means that when gluons are put
on-shell, they are not radiated in a forward cone around a heavy quark. This suppression of the available
phase space for radiation, the {\it dead-cone} effect, implies less energy loss for heavier quarks \cite{deadcone}.

\begin{figure}[t]
\begin{center}
\epsfig{file=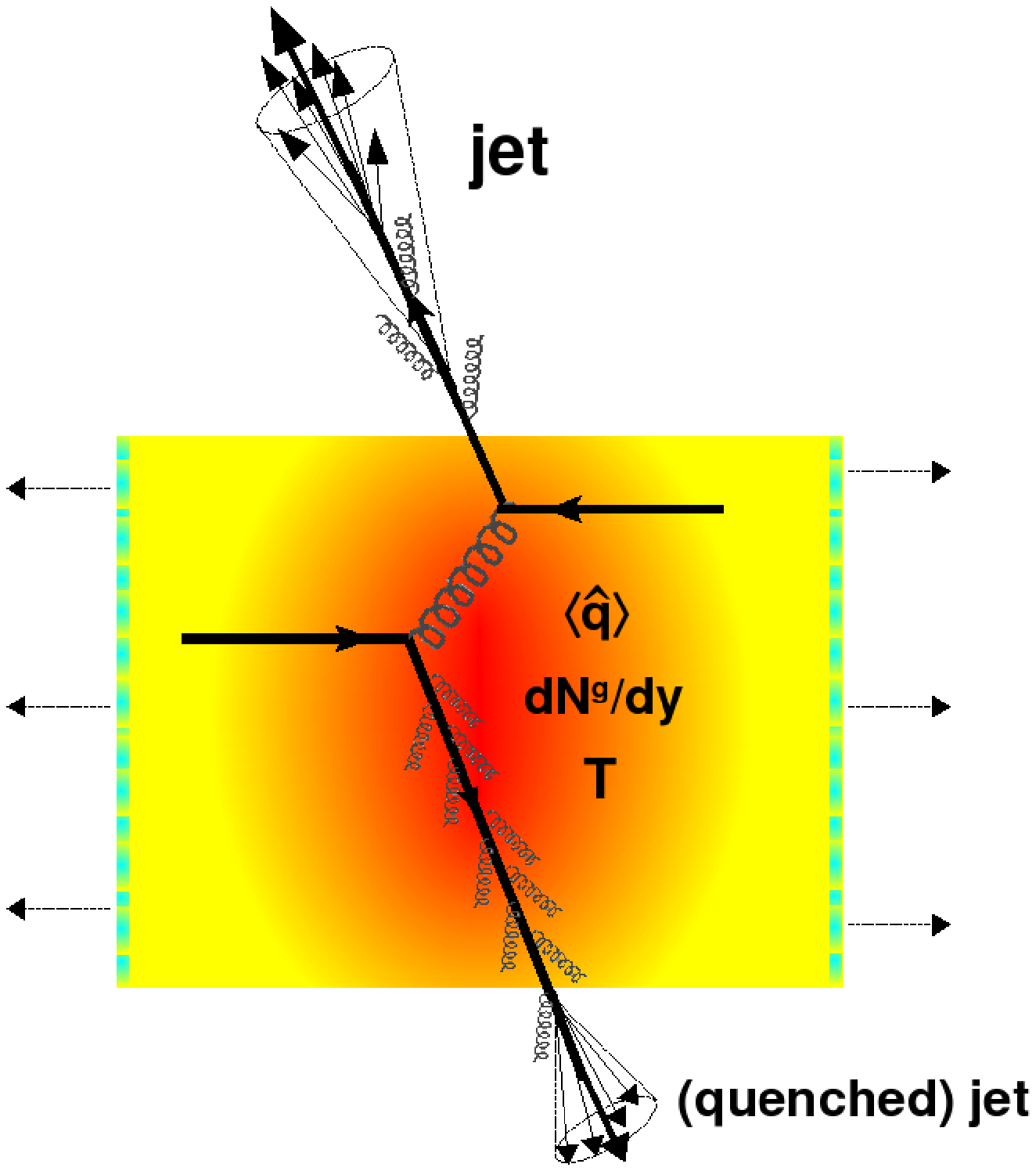,width=4.5cm}
\hspace{1cm}
\epsfig{file=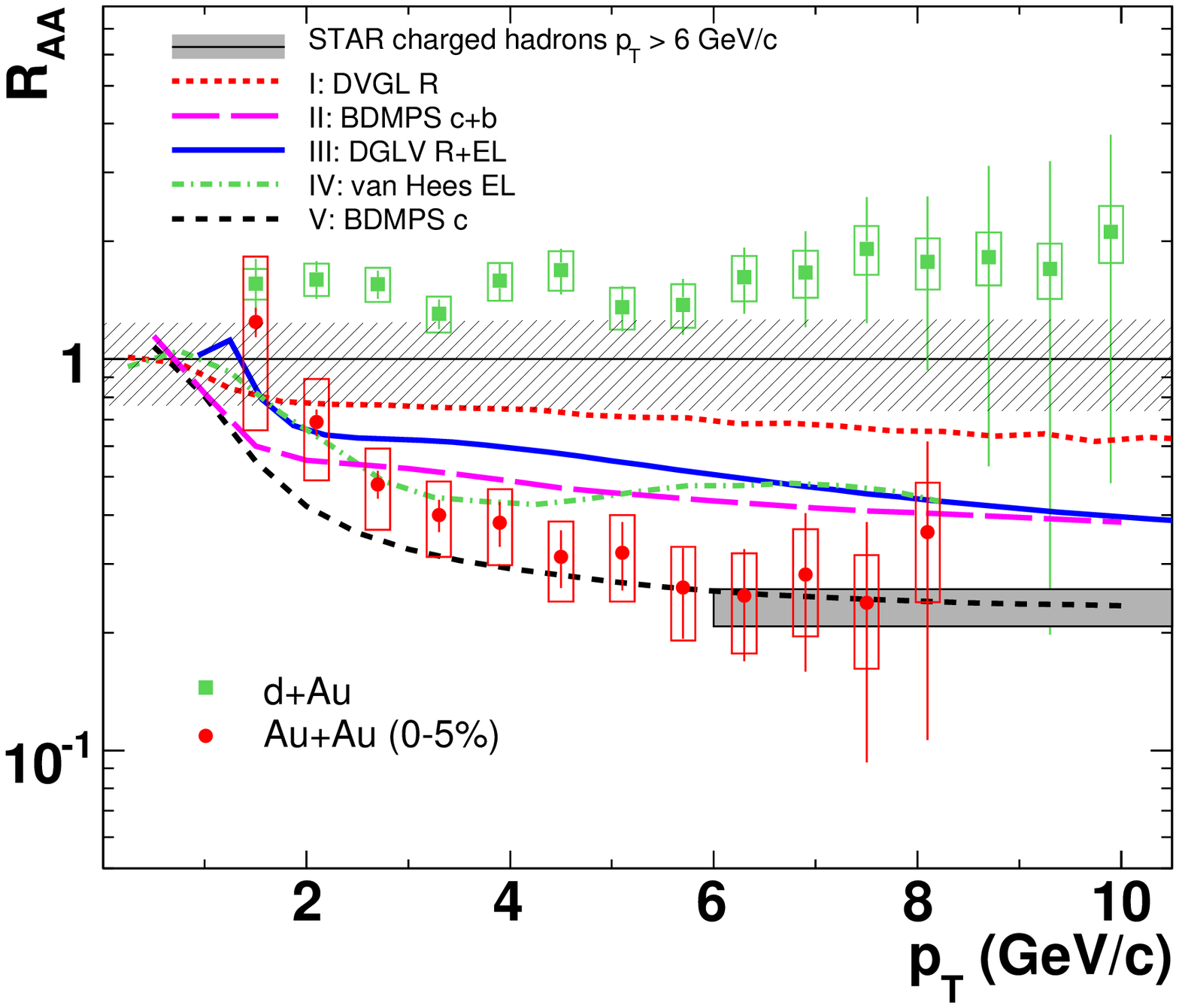,width=6.5cm}
\caption{Left: production of high-energy partons in a hard process, which then lose energy propagating through the plasma. Some quantum fluctuations in their wave function are put on shell while interacting with the medium and become emitted radiation.
Right: the resulting particle production in AA collisions is suppressed ($R_{AA}<1$) compared to independent nucleon-nucleon collisions. The suppression is large for light hadrons, and similar for heavy mesons (those data are displayed in the figure), which is difficult to accommodate in a weakly-coupled QCD description.}
\end{center}
\end{figure}

In pQCD, medium-induced gluon radiation is due to multiple scatterings of the virtual gluons.
If, while undergoing multiple scattering, the virtual gluons pick up enough transverse momentum to be put on shell,
they become emitted radiation. The accumulated transverse momentum squared picked up by a gluon of coherence
time $t_c$ is
\begin{equation}
p_\perp^2=\mu^2 \frac{t_c}{l}\equiv\hat{q}\ t_c
\end{equation}
where $\mu^2$ is the average transverse momentum squared picked up in each scattering, and
$l$ is the mean free path. These medium properties are involved through the ratio
$\hat{q}=\mu^2/l.$

Since only the fluctuations which pick up enough transverse momentum are freed ($k_\perp<p_\perp$),
the limiting value can be obtained by equating $k_\perp^2$ with
$p_\perp^2=\hat{q}\omega/k_\perp^2:$
\begin{equation}
k_\perp<(\hat{q}\omega)^{1/4}\equiv Q_s(\omega)\ .
\end{equation}
The picture is that highly virtual fluctuations with $k_\perp>Q_s$ do not have time to pick up enough $p_\perp$ to be freed, while the longer-lived ones with $k_\perp<Q_s$ do. That transverse momentum $Q_s$ which controls which gluons are freed and which are not is called the saturation scale. With heavy quarks, one sees that due to the dead cone effect, the maximum energy a radiated gluon can have is $\omega=\gamma k_\perp=\gamma Q_s$ (and its coherence time is $t_c=\gamma/Q_s)$. This allows to
estimate the heavy-quark energy loss:
\begin{equation}
-\frac{dE}{dt}\propto\alpha_sN_c\frac{\gamma Q_s}{\gamma/Q_s}=\alpha_s N_c Q_s^2\ .
\label{eloss}\end{equation}
The saturation momentum in this formula is the one that corresponds to the fluctuation which dominates the energy loss: $Q_s=(\hat{q}\gamma)^{1/3}.$

For a plasma of extend $L\!<\!t_c=\gamma^{2/3}/\hat{q}^{1/3},$ formula (\ref{eloss}) still holds but with $Q_s^2=\hat{q}L.$ These are the basic ingredients of more involved phenomenological calculations, but after comparisons with data, it has remained unclear if this perturbative approach can describe the suppression of high$-p_\perp$ particles. For instance, at RHIC temperatures, the value $\hat{q}\sim 1-3\ \mbox{GeV}^2/\mbox{fm}$ is more natural than the $5-10\ \mbox{GeV}^2/\mbox{fm}$ needed to describe the data on light hadron production. If one accepts to adjust $\hat{q}$ to this large value, then the $D$ and $B$ mesons are naturally predicted to be less suppressed than light hadrons, which is not the case (see Fig.4).

While the present pQCD calculations should still be improved, and may be shown to work in the future, this motivated to think about strongly-coupled plasmas. The tools to address the strong-coupling dynamics in QCD are quite limited, however for the $N=4$ Super-Yang-Mills (SYM) theory, the AdS/CFT correspondence is a powerful approach used in many studies. The findings for the strongly-coupled SYM plasma may provide insight for gauge theories in general, and some aspects may even be universal. One interesting result is that the total energy loss of hard probes goes as $\Delta E\propto L^3$ at strong coupling \cite{us}, instead of the $L^2$ law at weak coupling.

\begin{footnotesize}
\bibliographystyle{mpi08} 
{\raggedright
\bibliography{marquet}

\providecommand{\etal}{et al.\xspace}
\providecommand{\href}[2]{#2}
\providecommand{\coll}{Coll.}
\catcode`\@=11
\def\@bibitem#1{%
\ifmc@bstsupport
  \mc@iftail{#1}%
    {;\newline\ignorespaces}%
    {\ifmc@first\else.\fi\orig@bibitem{#1}}
  \mc@firstfalse
\else
  \mc@iftail{#1}%
    {\ignorespaces}%
    {\orig@bibitem{#1}}%
\fi}%
\catcode`\@=12
\begin{mcbibliography}{10}

\bibitem{cgcrev}
E.~Iancu and R.~Venugopalan~(2003).
\newblock \href{http://www.arXiv.org/abs/hep-ph/0303204}{{\tt
  hep-ph/0303204}}\relax
\relax
\bibitem{jyrev}
J.~Jalilian-Marian and Y.~V. Kovchegov,
\newblock Prog. Part. Nucl. Phys.{} {\bf 56},~104~(2006).
\newblock \href{http://www.arXiv.org/abs/hep-ph/0505052}{{\tt
  hep-ph/0505052}}\relax
\relax
\bibitem{glasma}
T.~Lappi and L.~McLerran,
\newblock Nucl. Phys.{} {\bf A772},~200~(2006).
\newblock \href{http://www.arXiv.org/abs/hep-ph/0602189}{{\tt
  hep-ph/0602189}}\relax
\relax
\bibitem{paulraju}
P.~Romatschke and R.~Venugopalan,
\newblock Phys. Rev.{} {\bf D74},~045011~(2006).
\newblock \href{http://www.arXiv.org/abs/hep-ph/0605045}{{\tt
  hep-ph/0605045}}\relax
\relax
\bibitem{hirokazu}
H.~Fujii and K.~Itakura,
\newblock Nucl. Phys.{} {\bf A809},~88~(2008).
\newblock \href{http://www.arXiv.org/abs/0803.0410}{{\tt 0803.0410}}\relax
\relax
\bibitem{glv1}
F.~Gelis, T.~Lappi, and R.~Venugopalan,
\newblock Phys. Rev.{} {\bf D78},~054019~(2008).
\newblock \href{http://www.arXiv.org/abs/0804.2630}{{\tt 0804.2630}}\relax
\relax
\bibitem{glv2}
F.~Gelis, T.~Lappi, and R.~Venugopalan,
\newblock Phys. Rev.{} {\bf D78},~054020~(2008).
\newblock \href{http://www.arXiv.org/abs/0807.1306}{{\tt 0807.1306}}\relax
\relax
\bibitem{javier}
J.~L. Albacete,
\newblock Phys. Rev. Lett.{} {\bf 99},~262301~(2007).
\newblock \href{http://www.arXiv.org/abs/0707.2545}{{\tt 0707.2545}}\relax
\relax
\bibitem{tuomas}
T.~Lappi,
\newblock J. Phys.{} {\bf G35},~104052~(2008).
\newblock \href{http://www.arXiv.org/abs/0804.2338}{{\tt 0804.2338}}\relax
\relax
\bibitem{reviewKW}
A.~Kovner and U.~A. Wiedemann~(2003).
\newblock \href{http://www.arXiv.org/abs/hep-ph/0304151}{{\tt
  hep-ph/0304151}}\relax
\relax
\bibitem{reviewdde}
D.~d'Enterria~(2009).
\newblock \href{http://www.arXiv.org/abs/0902.2011}{{\tt 0902.2011}}\relax
\relax
\bibitem{deadcone}
Y.~L. Dokshitzer and D.~E. Kharzeev,
\newblock Phys. Lett.{} {\bf B519},~199~(2001).
\newblock \href{http://www.arXiv.org/abs/hep-ph/0106202}{{\tt
  hep-ph/0106202}}\relax
\relax
\bibitem{us}
F.~Dominguez, C.~Marquet, A.~H. Mueller, B.~Wu, and B.-W. Xiao,
\newblock Nucl. Phys.{} {\bf A811},~197~(2008).
\newblock \href{http://www.arXiv.org/abs/0803.3234}{{\tt 0803.3234}}\relax
\relax
\end{mcbibliography}
}
\end{footnotesize}
\end{document}